\documentclass[conference]{IEEEtran}
\IEEEoverridecommandlockouts
\usepackage{cite}
\usepackage{amsmath,amssymb,amsfonts}
\usepackage{algorithmic}
\usepackage{graphicx}
\usepackage{textcomp}
\usepackage{xcolor}
\usepackage{fancyhdr}
\usepackage{multirow}
\def\BibTeX{{\rm B\kern-.05em{\sc i\kern-.025em b}\kern-.08em
    T\kern-.1667em\lower.7ex\hbox{E}\kern-.125emX}}
    
\begin{document}

\title{ \textbf{HOAA:} \textbf{H}ybrid \textbf{O}verestimating \textbf{A}pproximate \textbf{A}dder for Enhanced Performance Processing Engine\\
\thanks{This work is supported under the Special Manpower Development Program for Chip to Startup (SMDP-C2S), Ministry of Electronics and Information Technology (MeitY), Govt. of India. The authors are with the Department of Electrical Engineering, Indian Institute of Technology (IIT) Indore, Simrol-453552, India. (Corresponding author: Santosh Kumar Vishvakarma, \textbf{E-mail:} skvishvakarma@iiti.ac.in).}
}

\author{
    \IEEEauthorblockN{Omkar Kokane, Prabhat Sati, Mukul Lokhande, Santosh Kumar Vishvakarma}
    \IEEEauthorblockA{
        \textit{Department of Electrical Engineering} \\
        \textit{Indian Institute of Technology Indore}\\
        Indore, India \\
      $<$mt2302102023, ee230002052, phd2201102020, skvishvakarma$>$@iiti.ac.in
    }
}

\maketitle
\thispagestyle{fancy}

\begin{abstract}
This paper presents the Hybrid Overestimating Approximate Adder designed to enhance the performance in processing engines, specifically focused on edge-AI applications.
A novel Plus One Adder design is proposed as an incremental adder in the RCA chain, incorporating a Full Adder with an excess-1 alongside inputs A, B, and Cin. The design approximates outputs to 2-bit values to reduce hardware complexity and improve resource efficiency. The Plus One Adder is integrated into a dynamically reconfigurable HOAA, allowing runtime interchangeability between accurate and approximate overestimation modes. The proposed design is demonstrated for multiple applications, such as Two's complement subtraction and Rounding-to-even, and the Configurable Activation function, which are critical components of the Processing engine. Our approach shows a 21\% improvement in area efficiency and a 33\% reduction in power consumption, compared to state-of-the-art designs with minimal accuracy loss. Thus, the proposed HOAA could be a promising solution for resource-constrained environments, offering ideal trade-offs between hardware efficiency vs computational accuracy. 
\end{abstract}


\section{Introduction}
Deep Neural networks (DNN) have become ubiquitous in edge-AI devices, powering applications like facial recognition, object detection in computer vision, chatbots for human-like interactions, sentiment analysis in online shopping, social media, etc. In recent years, the models have grown exponentially in size and complexity\cite{a1}, contributing to improved prediction accuracy and effectiveness, potentially raising significant concerns for hardware platforms that are used for real-time inference. Software-level techniques like quantization and pruning help to provide smaller models for smooth inference\cite{a2}. However, to provide the best user experience, the need for hardware accelerators has rapidly evolved for resource-constrained environments\cite{a2,a4}. The edge accelerator must have special computing blocks to efficiently process various Convolution, Pooling, and Fully connected Layers. At the circuit level, the fundamental parallel components are Multiply-and-Accumulate (MAC) \cite{a3} operations and Non-linear Activation Functions (AF) \cite{a4}. However, as the main processor offloads these tasks to the accelerator, the need to support less frequent operations like Subtraction, Rounding, and Pooling operations has also arisen. Otherwise, these smaller workloads would contribute to more power consumption as these operations would need to be processed by a general-purpose processor and involve significant data movement. Understanding the trade-off between the NN performance vs physical parameters of hardware resources aids in accelerating the efficient acceleration of DNN solutions. Researchers\cite{a5,a6,a7} have explored architectural optimization at various abstraction levels, including fixed-point (FxP) arithmetic, hardware reuse/reduction, and approximation.  

\begin{table}[!t]
\caption{State-of-the-art approximate adders and their implications on neural network inference accuracy}
\label{sota-comp}
\renewcommand{\arraystretch}{1.5}
\scalebox{0.8}{
\begin{tabular}{|c|c|c|c|c|}
\hline
\textbf{Adder Unit} & \textbf{\# of gates} & \textbf{\begin{tabular}[c]{@{}c@{}}Hardware Efficiency\\ (Area-Power-Delay)\end{tabular}} & \textbf{\% Improvement} & \textbf{\begin{tabular}[c]{@{}c@{}}Accuracy\\ Overhead(\%)\end{tabular}} \\ \hline
\textbf{FA} & 40 & N/A & N/A & 0 \\ \hline
\textbf{HADD} & 32 & Area, Power & 50 & -1.5 \\ \hline
\textbf{LOA} & 25 & Area, Power, Delay & 60 & -3 \\ \hline
\textbf{ACA} & 32 & Power & 45 & 0 \\ \hline
\textbf{AMA} & 20 & Power & 70 & N/A \\ \hline
\end{tabular}}

\vspace{-5mm}
\end{table}

The article has discussed one fundamental digital block binary adder. The adders are widely adapted in ALUs (addition, subtraction) in microprocessors, Shift-Add-based MAC units, Coordinate Rotation Digital Computer (CORDIC) based transcendental functions, etc. The conventional accurate adders\cite{a5,a6} involve Ripple Carry Adders (RCA), Carry Look-Ahead Adders (CLA), Carry Bypass Adders (CBA), Carry Select Adders (CSA), Tree Adders, Serial Adders, Brent Kung Adders (BKA), Kogge Stone Adders (KSA), etc. However, The last decade has seen the rise of approximate adders by modifying the carry chain, half adder (HA), full adder (FA), compressor or reducing logic gates. The authors\cite{a6,a7} discussed the Area-Precision, Power-Precision, and Delay-Precision trade-off points for some of the approximate adders such as Almost Correct Adder (ACA-I/II), Quality-area optimal low-latency approximate Adder (QuAd), Equal Segmentation Adder (ESA), Error-Reduced Carry Prediction Approximate full Adder (ERCPAA), Generic Accuracy configurable adder (GeAr), Area-Power Efficient approximate adder (APEx), Error-Tolerant Adder (ETA-I/II), Lowest-cost Imprecise Adders (LIA) with error induced in the diverse application scenarios. ACA-I, ACA-II, and QuAd are designed to exploit the area-accuracy tradeoff by targeting less significant parts of an adder in an error-controlled manner. ESA, ERCPAA, and ETA-II focus on the decreasing critical path with less accurate computations while maintaining accuracy. ESA, GeAr allows the processing of additions in equal segments, configurable as per application requirements, and illustrates the basic principle behind HOAA concept. APEx, LIA focuses on the reduction of power for applications where exact results are not critical. Considering the error resiliency of neural networks, the techniques mentioned above may not affect output accuracy up to a certain point; however, beyond this threshold, enhancing physical design parameters often involves sacrificing accuracy, highlighting the importance of evaluating ideal trade-offs. 

The article proposes an n-bit Hybrid Overestimating Approximate Adder to simplify advanced arithmetic operations with minor area overhead. The principal contributions of this work are :

\begin{enumerate}
    \item \textbf{Resource efficient Plus One Adder (P1A):} An incremental approximate adder is proposed that can be incorporated into the RCA chain, which provides \\excess-1 output. The design approximates certain outputs to reduce hardware overhead; Hence, corresponding error metrics are also evaluated.

    \item \textbf{Dynamically Re-configurable HOAA:} HOAA adder architecture is proposed to be capable of runtime interchangeability by replacing the m-LSB FAs of RCA with P1A. This saves one clock cycle and supports multiple arithmetic operations with the same hardware blocks.

    \item \textbf{Evaluations for versatile test cases:} The proposed design is evaluated on Two's complement subtraction, Rounding-to-even and Configurable AF. Monte Carlo-based error distance calculations are provided. The ASIC physical design performance parameters are evaluated at \texttt{CMOS 28nm}.
\end{enumerate}

The rest of this paper is organized as follows. Section 2 gives the background and motivation behind the proposed P1A and HOAA. A detailed analysis of the proposed architecture and evaluation methodology is described in Section 3. The design trade-offs between error metrics and physical design parameters are discussed in Section 4. Section 5 briefly summarises the overall work and provides future directions.

\begin{figure}[t]
    \centering
    \includegraphics[width=0.5\textwidth]{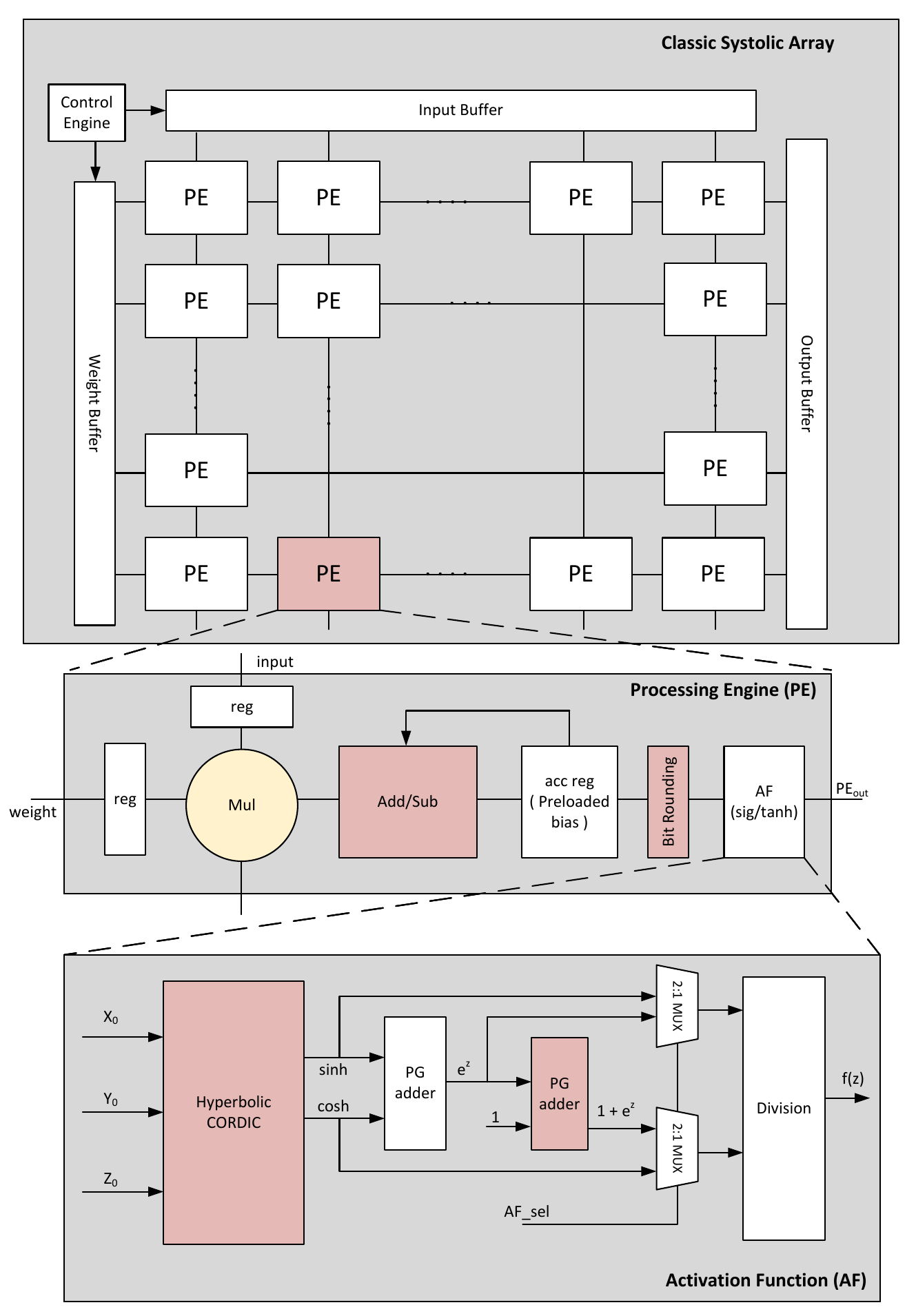} 
    \caption{Classic 2D systolic array architecture with PE highlighting the regions to be improved }
    \label{HOAA-SA}
    \vspace{-5 mm}
\end{figure}

\section{Background and Motivation}

There has been a sudden rise in the development of edge AI chips due to the high demand for running complex models and the need for optimized GEMM operations, which dominate the inference time. For instance, Google's Tensor Processing Unit (TPU) comprises 64K MAC units, and Samsung's Neural Processing Unit (NPU) has 6K MAC units\cite{a1,a7}. The DNN workloads LeNet, AlexNet, ResNet-50, and VGG-16 require 0.34M, 724M, 3.9B, and 15.5B MAC units \cite{a1,a2,a4}. The processing engine (PE) is the basic compute unit in a Convolutional Neural Network (CNN) accelerator, contributing to 90\% of energy efficiency. The PE\cite{a11,a12} comprises a multiplier followed by an adder, combined as MAC and AFs like ReLu, Sigmoid/Tanh. The fundamental block in the shift-add/sub-based CORDIC implementation\cite{a4} of PE is an adder. The equation \texttt{\ref{FA}} calculates the Sum and Carry in conventional FA.   

\begin{equation}
\begin{aligned}
\label{FA}
    \text{Sum} & = A \oplus B \oplus C_{\text{in}} \\
    \text{Cout} & = (A \cdot B) + (C_{\text{in}} \cdot (A \oplus B))
\end{aligned}
\end{equation}

\begin{figure*}[!t]
    \centering
    \includegraphics[width=\textwidth]{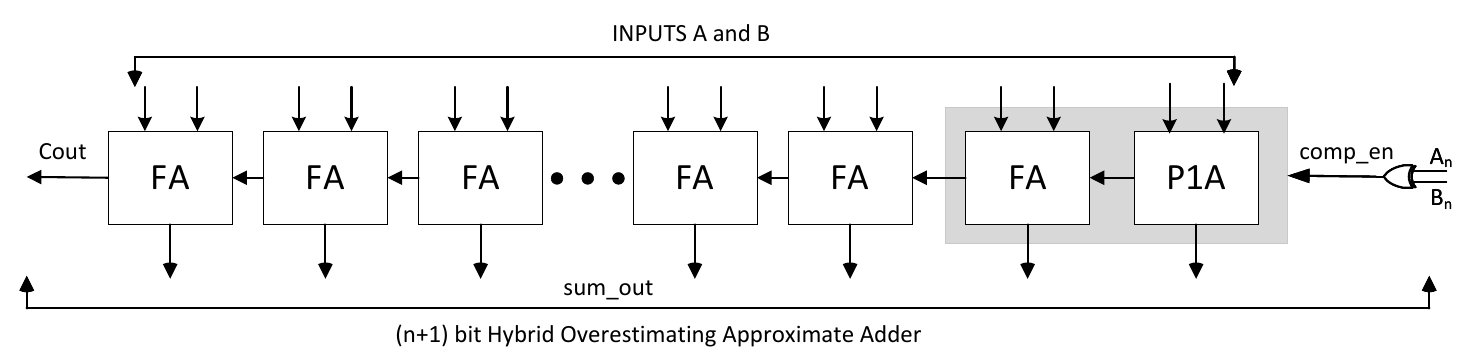} 
    \vspace{-5mm}
    \caption{N - Bit Re-Configurable Adder, (n+1) bit HOAA}
    \label{hybrid_adder}
        \vspace{-5 mm}
\end{figure*}

Various Adders have been proposed with different FA and HA combinations to fit the appropriate power, performance, and area trade-offs as shown in Table. \ref{sota-comp} according to various application scenarios. RCAs involve sequential carry propagation of parallel carry determination in CLA \cite{a4}. CBA selectively bypasses critical carry to optimize delay while CSA uses MUX to parallelize carry, showcasing area-speed tradeoff. While serial adders are suitable for low-power applications, Tree adders support distributed arithmetic with a hierarchical structure. BKA and KSA leverage parallel prefix carry computations suitable for high-performance, scalable applications.
Further, Given the inherent error-tolerant nature of CNNs, approximate computing methods have been explored widely for high hardware efficiency in implementing accelerators. 
Approximation is a vital element in enhancing physical design parameters and simplifying complex architecture by exploiting natural irregularities in real-time systems. The approach addresses resource management, with efficient circuits and helps for achieving faster and reduced computational demands. The design of approximate circuits involves using Majority-logic, modifying or removing logic gates from accurate circuits, or simplifying the Karnaugh map (K-map). From the initial application of approximate adders to boost the clock frequency of microprocessors in 2004, the AC applications have been automated with various innovative designs like the almost correct adder (ACA), the error-tolerant adder (ETA), the equal segmentation adder (ESA), and the approximate mirror adder (AMA), etc. The inherent redundancy in neuron computations and minimal degradation due to approximation errors open opportunities for trade-offs in energy efficiency, and compute density, making them ideal for resource-constrained environments like edge-AI devices. Additionally, the use of approximate adders can improve speed and throughput, leading to faster response time. Though beyond a certain error threshold, the performance of neural networks degrades noticeably, the approximation shall be tuned for efficient inference across diverse applications. A hybrid combination of conventional adders for MSBs and approximate adders for LSBs ensures minimum accuracy degradation (less than 5\%) with the equations provided in \texttt{\ref{HADD}}. This also ensures significant hardware efficiency in multi-bit fixed-point precision adders.

\begin{equation}
\begin{aligned}
\label{HADD}
    \text{Sum} & = (A + \text{Cin}) \oplus B \\
    \text{Carry} & = (A + \text{Cin}) \cdot B
\end{aligned}
\end{equation}

PE has been pivotal for the efficiency of the DNN accelerator as it contributes to 90\% operations. Despite huge efforts put in to enhance the performance, the PE encounters major performance fallback linked with the $+1$ operation. In \textbf{Signed MAC units}, $+1$ operation is crucial during Two's complement subtraction. In the quantization, \textbf{Rounding to even} technique employed within PE requires $+1$, contributing to one cycle delay. \textbf{Configurable Activation Function } with CORDIC methodology involves frequent $+1$ operation affecting the performance of overall PE and, subsequently, the entire systolic array DNN accelerator. The overhead would typically be the delay of one clock cycle in serial adders or area overhead in parallel adders, along with huge power consumption, depending on the bit precision of operands involved. An approximate P1A has been proposed to address the limitation, which performs the $+1$ operation within the same processing cycle. Therefore, the associated area or latency could be reduced by trading off slight accuracy in the runtime. We further analyzed the impact as well. Further dynamically reconfigurable P1A within the RCA chain has been proposed as novel HOAA. Hybrid approach-based HOAA ensures the runtime reconfigurability between FA and P1A enabling future scope for OEA. Different bit-width segments can be utilized with an approximate multiplier based on an error-controllable mechanism, fundamentally improving accuracy loss due to logarithmic MAC operations.

\section{Proposed Architecture}

We discuss the implementation of a P1A adder and extend it to HOAA architecture. We also evaluated the error analysis for the use cases with corresponding error metrics to demonstrate resource-efficient design. 

\subsection{Proposed P1A Design}

Signed MAC operations use an RCA chain with FAs in the accumulator stage. Tradition FA \textit{\ref{FA}} uses five logic gates are used to compute the sum and carry. However, reusing the adder as Two's complement-based subtraction typically consumes two-cycle latency and power consumption for N-bit operation. We identified the necessity of a $+1$ adder to address the issue and proposed the circuitry based on a modified truth table \textit{\ref{truthtable}} with the Karnaugh Map technique. Further, to obtain the approximate P1A, we evaluated the error metrics and corresponding hardware costs. The accurate $+1$ adder and approximate P1A equations are shown in Eq. \textit{\ref{AccP1A}}, \textit{\ref{AppP1A}} respectively. Compared to FA, the proposed P1A uses only three logic gates, highlighting hardware efficiency. The incorrect outputs are highlighted in Table \textit{\ref{truthtable}}. The truth table shows two errors in sum and carry-bit. Further, the error rate analysis is provided to justify the proposed approximations. 

\begin{equation}
    \begin{aligned}
     \centering
     \label{AccP1A}
\text{Sum} &= A \cdot C_{in} + B \cdot A + B \cdot C_{in} + \overline{A} \cdot \overline{B} \cdot \overline{C_{in}}, \\
        \text{Carry} &= A + B + C_{in}.
    \end{aligned}
\end{equation}

The circuit modifies with circuitry around to provide two modes with the same HOAA(N,m) to support both operations. The FAs are utilized during normal addition while keeping the P1A block disabled with PG for normal addition. Conventional N-bit RCA was modified typically by replacing the LSB adder with P1A for subtraction operation as illustrated in Fig. \textit{\ref{hybrid_adder}}, which also ensures the direct addition of excess-1 to LSB position and faster subtraction operation within the same cycles as that of addition. Notably, with increased adder size, the error introduced by P1A vanishes, approaching better suited for scalable architectures with minimal static power consumption.

\begin{table}[!t]
\centering
\caption{Truth table for accurate and approximate P1A adder designs, including all input combinations, output sum, and carry bits}
\renewcommand{\arraystretch}{1.25} 
\setlength{\tabcolsep}{6pt} 
\label{truthtable}
\begin{tabular}{|c|c|c|ccc|cc|}
\hline
 &  &  & \multicolumn{3}{c|}{Accurate P1A Output} & \multicolumn{2}{c|}{   Approx. P1A Output  } \\ \cline{4-8} 
\multirow{-2}{*}{A} & \multirow{-2}{*}{B} & \multirow{-2}{*}{Cin} & \multicolumn{1}{c|}{Sum} & \multicolumn{1}{c|}{Cout} & Cout2 & \multicolumn{1}{c|}{Sum} & \multicolumn{1}{c|}{Cout}  \\ \hline
0 & 0 & 0 & \multicolumn{1}{c|}{1} & \multicolumn{1}{c|}{0} & 0 & \multicolumn{1}{c|}{1} & 0 \\ \hline
0 & 0 & 1 & \multicolumn{1}{c|}{0} & \multicolumn{1}{c|}{1} & 0 & \multicolumn{1}{c|}{0} & 1 \\ \hline
0 & 1 & 0 & \multicolumn{1}{c|}{0} & \multicolumn{1}{c|}{1} & 0 & \multicolumn{1}{c|}{0} & 1 \\ \hline
0 & 1 & 1 & \multicolumn{1}{c|}{1} & \multicolumn{1}{c|}{1} & 0 & \multicolumn{1}{c|}{1} & 1 \\ \hline
1 & 0 & 0 & \multicolumn{1}{c|}{0} & \multicolumn{1}{c|}{1} & 0 & \multicolumn{1}{c|}{{\color[HTML]{FD6864} 1*}} & {\color[HTML]{FD6864} 0*} \\ \hline
1 & 0 & 1 & \multicolumn{1}{c|}{1} & \multicolumn{1}{c|}{1} & 0 & \multicolumn{1}{c|}{1} & 1 \\ \hline
1 & 1 & 0 & \multicolumn{1}{c|}{1} & \multicolumn{1}{c|}{1} & 0 & \multicolumn{1}{c|}{1} & 1 \\ \hline
1 & 1 & 1 & \multicolumn{1}{c|}{0} & \multicolumn{1}{c|}{0} & 1 & \multicolumn{1}{c|}{{\color[HTML]{FD6864} 1*}} & {\color[HTML]{FD6864} 1*} \\ \hline
\end{tabular}
    \vspace{-5 mm}
\end{table}

\begin{equation}
\begin{aligned}
     \label{AppP1A}
    Sum= A + \overline{B \oplus Cin}\\
    Cout = B    +   Cin
\end{aligned}
\end{equation}

\subsection{Proposed HOAA Design}

We evaluated two approaches to fit the proposed P1A into the adder design to support subtraction within the same processing cycle adaptively. The first approach was based upon Reconfigurable Approximate CLA, which would provide the reconfigurable ability to switch between approximate and accurate modes with MUX. The second approach involved using (N,m) adders as a combination of conventional and approximate adders, which has also been standard practice for hardware efficiency with minimal accuracy degradation. The runtime selection signal comp\_en has been generated based on MSBs of both operands for the selection between P1A and HA. This enables generalization and provides a template for a scalable architecture. Further, the approximate adders could be power gated when not required; however, this has not been evaluated in this work.

\begin{equation}
\begin{aligned}
\label{Timing}
    T_{sum}= T_{xnor}+T_{or}\\
    T_{carry}=T_{or}
    \end{aligned}
\end{equation}

Delay is one of the important parameters in PPA analysis. Even the SOTA studies highlight the importance of varied propagation delay per individual logic gates' properties. The propagation delay analysis for the sum and carry-bit of 1 bit P1A can be found in equations \textit{\ref{Timing}}. The further analysis of (N,m) HOAA follows the methodology per HADD/LOA adders. The equation \textit{\ref{AppP1A}} highlights linear area savings as the number of gates increases with bit-precision. Compared to the conventional adder, this would be an additional reduction in delay for any n-bit precision operation due to P1A.

\begin{figure}[!t]
    \centering
    \includegraphics[width=0.4\textwidth, height=0.085\textheight]{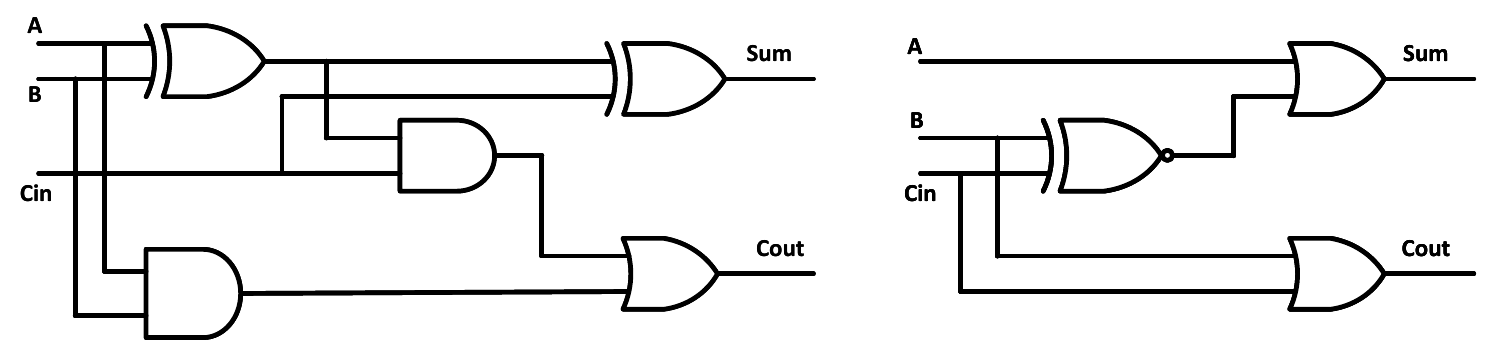} 
    \caption{1-bit Schematic of a) Conventional Full Adder b) Plus One Adder}
    \label{p1a_symbol}
        \vspace{-5 mm}
\end{figure}

\section{ Methodology \& Results Discussion}

The Conventional full adder (FA) truth table was modified by directly adding $+1$ to the {Cout, Sum} outputs while calculating the P1A equations. This modified truth table was then converted to Boolean equations using Karnaugh maps (K-maps), producing accurate equations \textit{\ref{AccP1A}}. However, the decision to introduce approximation was made due to the increased hardware cost of the equation. After evaluating numerous cases of the proposed addition and the corresponding accuracy loss, an approximate $+1$ adder was proposed, as per equation \textit{\ref{AppP1A}} and error analysis is mentioned in Table \textit{\ref{python_error}}. The timing analysis can be done for a single P1A as per equation \textit{\ref{Timing}}, replacing the combinational delay for a particular HA within RCA. Compared to conventional FA (28T), P1A requires only 16 transistors, significantly saving the CMOS area.

The proposed design has been validated with hardware-software co-design emulation flow for the experimental analysis. \texttt{Google Colab Notebook} with \texttt{Python v3.0} has been utilised for error metrics calculations as per \cite{a4, a6,a7}. The functional emulation for the proposed hardware architecture was done with the of \texttt{Python v3.0} and the error metrics were calculated as shown in Table \textit{\ref{python_error}}. The quantitative measurement of error introduced due to approximation, and its impact on performance, and quality of approximate neural networks, further to assist designers in understanding the more suitable trade-offs for particular applications, error metrics are essential. The error metrics to evaluate the accuracy loss incurred include Mean Square Error (MSE), Normalised Mean Error Distance (NMED), and Mean Relative Error Distance (MRED) \cite{a4,a6}. MSE provides overall error magnitude, as in the average squared difference between exact and approximate results. The MSE of less than 2\% shows the efficiency of design for all the test cases. NMED and MRED give a relative measure of error and indicate error scaling with architecture. MRED and NMED around 6\% show the comparison of errors across different scale and application scenarios. Further, The Verilog RTL code for the proposed Adder was written and functionally validated using \texttt{Xilinx Vivado 2023.2} for hardware implementation. 

During ASIC analysis, the 8-bit HOAA is synthesized with \texttt{Synopsys Design Compiler} using \texttt{CMOS 28nm PDK} at a power supply of 0.9V for the configurations shown in Table \textit{\ref{PD28}}. The physical parameters were evaluated at the \texttt{HPC+} node, considering the better compute density, reduced area-power-dealy, and industry-adopted solution for resource-constrained scenarios. Monte-Carlo simulations for uniformly distributed 2\textsuperscript{n+1} times random input pattern variable. For the calculation of error metrics \textit{\ref{python_error}}, the simulations were computed on random inputs and were compared with true outputs from \texttt{Python Numpy}. Approximation trade-offs were evaluated to check the accuracy and compare the adder architectures. The proposed adder signifies minimal accuracy loss compared to LOA while absolute gains in PPA design parameters relative to FA, AMA, HADD, and SESA-I. P1A shows an improvement of around 5\% in area compared to HADD\cite{a3} and AMA\cite{a10}, while consuming approximately 15\% less power than HADD\cite{a3} and SESA-I\cite{a8}. Additionally, P1A demonstrates a 21\% reduction in area and a 33\% reduction in power consumption compared to the conventional FA. The slack calculation is performed with a period of 10ns or an operating frequency of 100 MHz, and the value reflects a smaller improvement in timing margins compared to FA and hence, a higher operating frequency. 

\begin{table}[!t]
    \vspace{-5 mm}
\centering
\caption{Error metrics evaluation for 8-bit HOAA design
\\
$\ast$ All values are in percentage scale}
\setlength{\tabcolsep}{12pt}
\renewcommand{\arraystretch}{1.25}
\label{python_error}
\begin{tabular}{|c|c|c|c|}
\hline
\textbf{Error Metrics } & \textbf{Case-I} & \textbf{Case-II} & \textbf{Case-III} \\ \hline
\textbf{MSE} & 0.02444  & 0.02406 & -0.06766 \\ \hline
\textbf{NMED} & 0.02444  & 0.02406  & 0.06766  \\ \hline
\textbf{MRED} & 0.06834  & 0.06729  & 0.06759 \\ \hline

\end{tabular}

\end{table}

\begin{table}[!t]
\centering
\caption{Physical Design Parameters at CMOS 28 nm, VDD 0.9V, frequency 100 MHz}
\renewcommand{\arraystretch}{1.25} 
\setlength{\tabcolsep}{12pt} 
\begin{tabular}{|c|c|c|c|}
\hline
Attributes & Area ($\mu m^{2}$ ) & Power ($\mu W$) & Slack (ns) \\ \hline
FA & 8.736 & 1.164 & 1.87 \\ \hline
\begin{tabular}[c]{@{}c@{}}HADD\cite{a3}\end{tabular} & 7.392 & 0.649 & 1.91 \\ \hline
\begin{tabular}[c]{@{}c@{}}SESA-1\cite{a8}\end{tabular} & 6.384 & 0.921 & 1.93 \\ \hline
\begin{tabular}[c]{@{}c@{}}LOA\cite{a9}\end{tabular} & 4.032 & 0.567 & 1.98 \\ \hline
\begin{tabular}[c]{@{}c@{}}AMA\cite{a10}\end{tabular} & 6.552 & 0.810 & 1.93 \\ \hline
{\color[HTML]{00009B} \begin{tabular}[c]{@{}c@{}}P1A\end{tabular}} & {\color[HTML]{00009B} 6.888} & {\color[HTML]{00009B} 0.782} & {\color[HTML]{00009B} 1.93} \\ \hline
\end{tabular}

\label{PD28}
\vspace{-3 mm}
\end{table}

\subsection{Case-I: Subtraction}

Two's complement-based subtraction in signed computer arithmetic is the simplest and most efficient methodology followed. It simplifies subtraction between two binary numbers by modifying into the addition problem, which is convenient from a reconfigurable hardware implementation perspective and overflow handling. Negative numbers are represented with two's complement format by calculating one's complement/simply reverting the number, followed by adding one to LSB, which is a two-cycle process. The effective computational resources required for the process are two-cycle latency and power consumption of N-bit adders, where the second cycle is just utilized for adding "1". P1A can be effectively used to reduce the design complexity and enhance the throughput for such operations in Edge-AI applications. 

The differences between accurate and approximate computations have been measured with error metrics like MSE, NMED and  MRED and reported in the table \textit{\ref{python_error}}. It is also noteworthy that overall computational accuracy ain't much affected for 8-bit due to the introduction of P1A in LSB of HOAA and is reflected in overall better results. Hence, it can be concluded that the proposed HOAA can be a good choice for designing reconfigurable subtraction hardware.

\subsection{Case-II: Rounding-to-even}

In neural network computations, MAC units sum up the products of numbers, increasing precision. To support the limited precision requirement at Edge-AI, the numbers are often required to undergo quantization, resulting in rounding errors. Regular rounding accumulates the statistical bias due to long sequence operations in DNN, either in a positive direction or negative direction. Hence, the \texttt{IEEE754-2008} roundTiesToEven technique is more suitable as rounding errors are canceled due to adding \textit{1} when odd numbers are rounded down. However, the two-cycle process is more hardware-intensive due to the presence of shifters in two cycles, where one cycle is wasted simply for adding \textit{1}. This can be effectively optimized by introducing P1A, which results in the further reduction of computational resources. Our results from Table \textit{\ref{python_error}}, \textit{\ref{PD28}} highlight the energy efficiency of the design with minimal accuracy loss, even less than 1\% error. 

\begin{figure}
    \centering
    \includegraphics[width=0.4\textwidth, height=0.175\textheight]{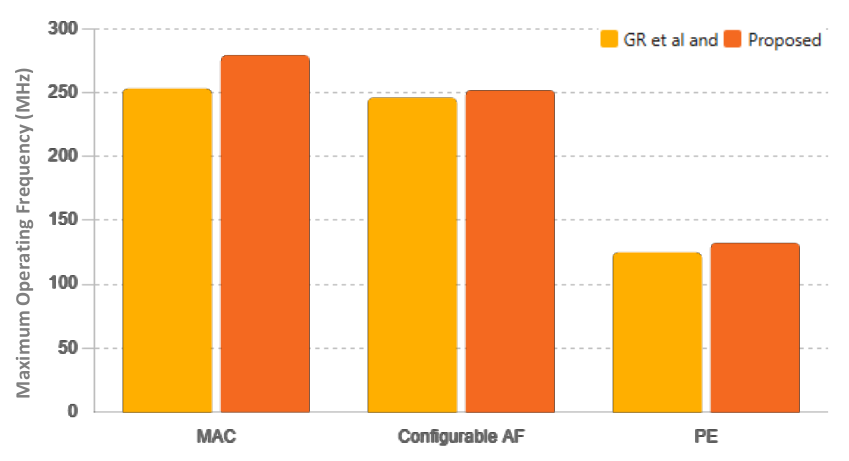}
    \caption{Comparison for Maximum Operating Frequency}
    \label{opcomp}
\end{figure}

\subsection{Case-III: Configurable activation function}

DNN inference models rely heavily on Non-linear AF to characterize input-output relationships. The hardware implementation methodologies for the implementation of reconfigurable activation functions are resource-intensive, and popular ones like \textit{sigmoid} and \textit{tanh} often require huge optimizations for Edge-AI applications. Iterative CORDIC-based methodology has been proven hardware resource-efficient yet affects critical delay. The runtime reconfigurable choice between \textit{sigmoid} and\textit{tanh} has been provided with AF\_sel signal. 
The CORDIC approach involves the computation of hyperbolic sine (\textit{sinh(Z)}) and cosine (\textit{cosh(Z)}), requiring two-stage adders before division operation to derive the \textit{sigmoid} function as shown in Eq. \textit{\ref{sigmoid}}. Our HOAA approach could easily enhance throughput by minimizing adding \textit{1}. Furthermore, Table \textit{\ref{PD28}} discusses the impact of approximation introduced by P1A, which is negligible. Case \textit{III} also demonstrates the scalability of our approach for fixed-point arithmetic, broadening P1A's applicability beyond the integer computations discussed earlier. 

\begin{equation}
\begin{aligned}
\label{sigmoid}
    e\textsuperscript{Z} = \cosh (Z) + \sinh (Z)\\
   \textit{sigmoid(Z)}= \frac{e\textsuperscript{Z}}{e\textsuperscript{Z} + 1} 
    \end{aligned}
\end{equation}

\vspace{0.5mm}

The count and placement of reconfigurable P1A in HOAA can be error-controlled while the decision to choose HOAA can be made on the evaluation of MSE, NMED, and MRED (Refer to Table \ref{python_error}) vs physical parameters (Refer to Table \ref{PD28}). However, while the error metrics are preliminary factors, a thorough analysis should be carried out for real-world DNN workloads. The proposed design shows significant improvement in the maximum operating frequency around 10\% for MAC while 5\% for AF and PE over \cite{a4,a11}. Thus, HOAA-incorporated PE is a viable solution for resource-constrained environments.

\vspace{0.5mm}
\section{Conclusion}

In this article, we proposed the design of P1A and integrated version HOAA. The novel adder bettered the Two's complement-based subtraction, Rounding to even, and configurable AF operations by performing $+1$ operation within the same processing cycle. The proposed adder is proven with mathematical analysis, the proposed hardware-software co-emulation is explained in the experimental results, and it was found to be better by 21\% area improvement and 33\% power-reduction when compared with SOTA designs. The analysis highlights minor accuracy loss, which is negligible compared to gains in hardware efficiency. Hence, the proposed HOAA design enables efficient PE design and is a promising solution for improved performance of edge-AI accelerators. The configurable P1A (HOAA) can be further extended in combination with underestimating approximate multipliers to enhance the accuracy of neural networks while maintaining the resource efficiency achieved through MAC approximation.

\end{document}